\documentclass[graybox, secnum]{svmult}

\usepackage{mathptmx}        
\usepackage{amsmath}
\usepackage{amssymb}
\usepackage{color}
\usepackage{helvet}          
\usepackage{courier}         
\usepackage{dirtree}

\usepackage{makeidx}         
\usepackage{graphicx}        
\usepackage{subfig}

\usepackage{multicol}        
\usepackage[bottom]{footmisc}
\usepackage{soul}            
\usepackage{hyperref}        
\hypersetup{colorlinks=true,urlcolor=blue}

\usepackage[misc]{ifsym}

\makeindex             

\usepackage{amsfonts}
\usepackage{latexsym}
\usepackage{mathrsfs}  
\usepackage{indentfirst}
\usepackage{cite}

\newcommand{\beq}{\begin{eqnarray}}
\newcommand{\eeq}{\end{eqnarray}}
\newcommand{\pa}{\partial}

\newcommand{\diag}{\,\mbox{diag}\,}

\newcommand{\eq}[1]{(\ref{#1})}
\newcommand{\n}[1]{\label{#1}}
\newcommand{\nn}{\nonumber}

\newcommand{\al}{\alpha}
\newcommand{\be}{\beta}
\newcommand\ga{\gamma}
\newcommand\Ga{\Gamma}

\newcommand\de{\delta}
\newcommand\De{\Delta}

\newcommand\ze{\zeta}

\renewcommand\th{\theta}

\newcommand\la{\lambda}

\newcommand\ph{\varphi}

\newcommand\na{\nabla}
\renewcommand\pa{\partial}

\newcommand{\rd}{\mathrm{d}}  

\DeclareSymbolFont{matha}{OML}{txmi}{m}{it}
\DeclareMathSymbol{\varv}{\mathord}{matha}{118}

\begin{document}


\title*{Black holes in non-local gravity}

\author{Luca Buoninfante, Breno L. Giacchini and Tib\'{e}rio de Paula Netto}
\institute{Luca Buoninfante \at Nordita, KTH Royal Institute of Technology and Stockholm University, Hannes Alfvéns v\"ag 12, Stockholm, SE-106 91, Sweden,
\email{luca.buoninfante@su.se}
\and
Breno L. Giacchini \at Department of Physics, Southern University of Science and Technology, Shenzhen 518055, China, \email{breno@sustech.edu.cn}
\and Tib\'{e}rio de Paula Netto \at Department of Physics, Southern University of Science and Technology, Shenzhen 518055, China \email{tiberio@sustech.edu.cn}}
%
%
\maketitle

\abstract{In this chapter we present a status report of black hole-like solutions in 
non-local theories of gravity in which the Lagrangians are at least quadratic in curvature and contain specific 
non-polynomial 
(\textit{i.e.}, non-local) operators. 
In the absence of exact black hole solutions valid in the whole spacetime, most of the literature on this topic focus on approximate and simplified equations of motion, which could provide insights on the full non-linear solutions. Therefore, the largest part of this chapter is devoted to the linear approximation. 
We present results on stationary metric solutions (including both static and rotating cases) and dynamical spacetimes describing the formation of non-rotating mini black holes by the collapse of null shells. 
Non-local effects can regularize the curvature singularities in both scenarios and, in the dynamical case, there exists a mass gap below which the formation of an apparent horizon can be avoided.
In the final part we discuss interesting attempts towards finding non-linear black hole solutions in non-local gravity.
Throughout this chapter, 
instead of focusing on a particular non-local model, we present results valid for large classes of theories (to a feasible extent). This more general approach allows the comparison of similarities and differences of the various types of non-local gravity models.
}

\section*{Keywords} 
Black holes, non-local gravity, ghost-free gravity, curvature singularity, regular solutions, non-locality, infinite derivatives


\section{Introduction}
\label{sec1}

Einstein's general relativity (GR) has been the most successful theory to describe classical aspects of gravity so far as many of its predictions have been confirmed to high precision~\cite{Will:2018bme}. Despite its great achievements, there are still unsolved problems suggesting that GR can only be used as an effective field theory of gravitational interaction, which works very well at low energy but breaks down in the ultraviolet (UV) regime. In fact, at the classical level GR is plagued by the presence of black-hole and cosmological singularities~\cite{Hawking:1973uf},  while at the quantum level the theory is perturbatively non-renormalizable~\cite{tHooft:1974toh,Goroff:1985sz}. 

A natural way to address these issues is to extend GR by adding higher-curvature terms to the Einstein--Hilbert action. Remarkably, a theory of gravity quadratic in curvature can be perturbatively renormalizable in four dimensions~\cite{Stelle77}. However, this higher-derivative theory turns out to be pathological, because of a massive spin-$2$ ghost-like degree of freedom that causes instabilities and breaks the $S$-matrix unitarity. 

This type of unhealthy degree of freedom can be avoided if the action is \emph{non-polynomial} in field derivatives. Indeed, by considering quadratic-curvature terms that contain specific infinite-derivative operators 
one can prevent the appearance of extra pathological modes in the physical spectrum, and still have an improved propagator in the UV, making it possible to formulate theories of gravity that are ghost-free and renormalizable~\cite{Krasnikov,Kuzmin,Tomboulis:1997gg,Biswas:2005qr,Modesto:2011kw,Biswas:2011ar}. The presence of non-polynomial differential operators renders the gravitational action \emph{non-local}.

In this chapter, we discuss black-hole solutions and alternatives in the context of 
non-local
theories of gravity that are UV-modifications of GR. 
The presentation has the form of a status report on
this topic and it reviews the most relevant results 
obtained so far, 
providing the essential details of both conceptual and computational aspects. 
We emphasize that, owing to the difficulties involved, a rigorous study of exact black-hole solutions in these models is still lacking. Therefore, the results compiled here mainly consist of approximate solutions 
valid in certain regimes, that might be useful to understand the behaviour of the full non-linear ones.

The chapter is organized as follows:
\begin{description}
	\item[\textbf{Sec.~\ref{quad-theories}:}] Since most of the considerations are carried out in the linear regime, in this section we briefly review the linearised formulation of ghost-free non-local theories of gravity, and show several interesting models that have been intensively studied in the literature.
	
	\item[\textbf{Sec.~\ref{stationary sol}:}] We discuss linearised stationary solutions in two situations, namely, static and rotating cases. We show that non-locality can regularize both point-like and ring-like singularities, and make several remarks about the physical implications. In what concerns the resolution of point-like singularities, we show that this regularization can occur at different levels (\textit{i.e.,} potential, curvature invariants and curvature-derivative invariants), depending on the type of form factor used.
	
	\item[\textbf{Sec.~\ref{mini-bh}:}] The dynamical solution of mini black hole formation by the collapse of null shells is discussed.  We consider the cases of thin and thick shells. In both cases there exists a mass gap below which the formation of an apparent horizon can be prevented, while the curvature singularities can also be regularized if the imploding shell has some non-vanishing thickness.
	
	\item[\textbf{Sec.~\ref{sec:non-linear}:}] We discuss the difficulties introduced by non-linearities in the field equations and review an interesting attempt towards understanding the effects of non-locality at the non-linear level. By working in a simplified situation, it is possible to obtain regular black hole solutions with a mass gap and with multiple horizons.
	
	\item[\textbf{Sec.~\ref{sec:conclus}:}] We present our concluding remarks.
\end{description}

Throughout this chapter we use the mostly plus metric convention, with the
Minkowski metric $\eta_{\mu\nu} = \diag (-1,+1,+1,+1)$.
The Riemann curvature tensor is 
defined by
\beq
\n{Rie}
R^\al {}_{\be\mu\nu} \,=\, 
\pa_\mu \Ga^\al_{\be \nu} - \pa_\nu \Ga^\al_{\be \mu} 
+ \Ga^ \al_{\mu \tau} \, \Ga^\tau_{\be \nu}  
- \Ga^\al_{\nu \tau } \, \Ga^\tau _{\be \mu} 
\, ,
\eeq
while 
$R_{\mu\nu} = R^\al {}_{\mu\al\nu}$ and $R = g^{\mu\nu} R_{\mu\nu}$ are, respectively, 
the Ricci tensor and the scalar curvature. 
Also, we adopt the unit system such that $c = 1$ and $\hslash  = 1$.


\section{Non-local theories of gravity}
\label{quad-theories}

Einstein's GR is described in terms of a very simple Lagrangian which is linear in curvature. A natural way of modifying the theory in the UV regime is to generalise the Einstein--Hilbert action through the inclusion of higher-order curvature invariants which contain higher derivatives. Local higher-derivative theories of gravity are usually considered pathological because of the ghost-like degrees of freedom that occur in the spectrum, in the conventional quantum field theory framework. However, as we discuss in this section, ghosts can be avoided if the principle of locality is given up.

Let us consider the generic action quadratic in curvature,
\begin{eqnarray}
S &=& \frac{1}{2\kappa^2}\int {\rm d}^4x\sqrt{-g} \, \Big\lbrace  R+\frac{1}{2} [ \, R \mathcal{F}_1(\Box)R+R_{\mu\nu} \mathcal{F}_2(\Box)R^{\mu\nu} \nonumber\\
&& \hspace{0.6cm} + R_{\mu\nu\rho\sigma}\mathcal{F}_3(\Box)R^{\mu\nu\rho\sigma} \, ] \Big\rbrace\,,\label{quad-action}
\end{eqnarray}
where $\kappa=\sqrt{8\pi G},$ $G$ is Newton's constant and the form factors $\mathcal{F}_i(\Box)$ are non-local, {\it i.e.}, non-polynomial functions of the d'Alembertian $\Box=g^{\mu\nu}\nabla_{\mu}\nabla_\nu.$ In particular, we assume that $\mathcal{F}_i(z)$ are analytic functions around $z=0$, in order to guarantee a smooth and consistent infrared limit. Therefore, expanding them in Taylor series,
\begin{eqnarray}
\mathcal{F}_i(\Box)=\sum\limits_{n=0}^{\infty} f_{i,n}\Box^n\,,\qquad i=1,2,3, 
\end{eqnarray}
and using the identity (see, {\it e.g.},~\cite{Deser:1986xr,AsoreyLopezShapiro})
\begin{equation}
	R_{\mu\nu\rho\sigma}\Box^n R^{\mu\nu\rho\sigma}=4 R_{\mu\nu}\Box^n R^{\mu\nu}- R\Box^n R+ O(R^3_{\ldots} ) + {\rm div}\,,
\end{equation}
the action~\eqref{quad-action} can be rewritten as 
\begin{equation}
	S= \frac{1}{2\kappa^2}\int \rd^4 x\sqrt{-g} \, \Big\lbrace  R + \frac{1}{2} \left[ \, R {F}_1(\Box) R+ R_{\mu\nu} {F}_2(\Box) R^{\mu\nu} + O(R^3_{\ldots} ) \, \right] \Big\rbrace  ,
	\label{quad-action-reduced}
\end{equation}
where 
\begin{equation}
	{F}_1(\Box) = \mathcal{F}_1(\Box)-\mathcal{F}_3(\Box)\,,\quad\quad  {F}_2(\Box)=\mathcal{F}_2(\Box)+4\mathcal{F}_3(\Box)\,, 
\end{equation}
and we have omitted boundary terms.

Since most part of this chapter is based on 
the linearised equations of motion, which come from the
second-order metric perturbations around the Minkowski spacetime at action level, we neglect the cubic-curvature terms $O(R^3_{\ldots} )$ in~\eq{quad-action-reduced} (or, in other words, the Riemann-squared term $ R_{\mu\nu\rho\sigma} {\cal F}_3(\Box) R^{\mu\nu\rho\sigma}$). Therefore, in what follows we analyse the gravitational action
\begin{equation}
	S= \frac{1}{2\kappa^2}\int \rd^4x \sqrt{-g}\, \Big\lbrace  R+\frac{1}{2} \left[ \, R {F}_1(\Box) R+ R_{\mu\nu} {F}_2(\Box) R^{\mu\nu} \, \right] \Big\rbrace \,.
	\label{quad-action-reduced-nocubic}
\end{equation}
%


\subsection{Linearised theory}\label{sec:lin-theory}

In the weak gravitational field approximation, we consider linear perturbations around the Minkowski background,
\begin{equation}
g_{\mu\nu}=\eta_{\mu\nu}+\kappa h_{\mu\nu}\,, 
\label{lin-metric} 
\end{equation}
and expand the action~\eqref{quad-action-reduced-nocubic} up to order $O(h^2_{\ldots})$~\cite{Modesto:2011kw,Biswas:2011ar,hedao}, namely,
\beq
S^{(2)}&=&  \frac{1}{4}\int {\rm d}^4x\, \Big\lbrace \frac{1}{2}h_{\mu\nu}a(\Box)\Box h^{\mu\nu}-h_{\mu}^{\sigma}a(\Box)\partial_{\sigma}\partial_{\nu}h^{\mu\nu}+h c(\Box)\partial_{\mu}\partial_{\nu}h^{\mu\nu} \nn\\
&& \vspace{1cm} -\frac{1}{2}h c(\Box)\Box h +\frac{1}{2}h^{\lambda\sigma}\frac{a(\Box)-c(\Box)}{\Box}\partial_{\lambda}\partial_{\sigma}\partial_{\mu}\partial_{\nu}h^{\mu\nu}\Big\rbrace \,,
\label{lin-quad-action}
\eeq
where 
\beq
		a(\Box) & \equiv &   1+\frac{1}{2}{F}_2(\Box)\Box,\\
		c(\Box) & \equiv &  1-2{F}_1(\Box)\Box-\frac{1}{2}{F}_2(\Box)\Box ,
\eeq
and $h\equiv\eta_{\mu\nu}h^{\mu\nu}$  is the trace of the field perturbation.
From Eq.~\eqref{lin-quad-action} we can derive the linearised field equations,
\beq	
		&& \hspace{-0.5cm} a(\Box)\left(\Box h_{\mu\nu}-\partial_{\sigma}\partial_{\nu}h_{\mu}^{\sigma}-\partial_{\sigma}\partial_{\mu}h_{\nu}^{\sigma}\right) \displaystyle +c(\Box)\left(\eta_{\mu\nu}\partial_{\rho}\partial_{\sigma}h^{\rho\sigma}+\partial_{\mu}\partial_{\nu}h-\eta_{\mu\nu}\Box h\right) \nonumber \\
		&& \hspace{2.2cm}  + \frac{a(\Box)-c(\Box)}{\Box}\partial_{\mu}\partial_{\nu}\partial_{\rho}\partial_{\sigma}h^{\rho\sigma}=-2\kappa T_{\mu\nu}\,,
		\label{lin-field-eq}
\eeq
where the coupling to matter is introduced through the energy-momentum tensor
\begin{equation}
	T_{\mu\nu}=-\frac{2}{\sqrt{-g}}\frac{\delta S_\text{m}}{\delta g^{\mu\nu}}\simeq \frac{2}{\kappa}\frac{\delta S_\text{m}}{\delta h^{\mu\nu}}\,.
	\n{EMT}
\end{equation}
In Eq.~\eq{EMT} $S_\text{m}$ is the matter action. The energy-momentum tensor satisfies the conservation law $\partial_{\mu}T^{\mu\nu}=0,$ consistently with the Bianchi identity. 


\subsection{Propagator}\label{sec:propag}

The propagator around the Minkowski background is computed by inverting the kinetic operator in the Lagrangian~\eqref{lin-quad-action}. Its gauge-fixing independent part reads~\cite{Tomboulis:1997gg,Modesto:2011kw,Biswas:2011ar,Accioly:2002tz}
\begin{equation} \label{Propagator}
\Pi_{\mu\nu\rho\sigma}(k)=\frac{{P}^{(2)}_{\mu\nu\rho\sigma}}{ k^2 f_2(-k^2)}-\frac{{P}^{(0-s)}_{\mu\nu\rho\sigma}}{2  k^2 f_0(-k^2)}\,,
\end{equation}
where 
\begin{equation} \label{f2-f0}
f_2(-k^2)\equiv a(-k^2)\,,\qquad f_0(-k^2)\equiv \frac{3c(-k^2)-a(-k^2)}{2}\,,
\end{equation}
and ${P}^{(2,0-s)}_{\mu\nu\rho\sigma}$ are the Barnes--Rivers operators that project the metric fluctuation $h_{\mu\nu}$ along its gauge-independent spin-$2$ and spin-$0$ components~\cite{Barnes,Rivers,VanNieuwenhuizen:1973fi}. 

In general, if the functions $f_{0,2}(-k^2)$ are polynomials (or, equivalently, if the form factors ${F}_i(\Box)$ are polynomials) the propagator~\eq{Propagator} possesses, besides the massless graviton, extra massive degrees of freedom that can cause instabilities. This happens because some of these modes are associated with real poles with negative residue, {\it i.e.}, they are unhealthy ghosts. 

However, in the framework of non-local models it is possible to construct classes of ghost-free theories by imposing some requirements on the form factors.
Indeed, if
\begin{eqnarray}
f_s(-k^2) = e^{\gamma_s(-k^2)}\,, \label{ghost-free-choice} \qquad s = 0,2\,,
\end{eqnarray}
where $\gamma_2(-k^2)$ and $\gamma_0(-k^2)$ are entire functions, the propagator only has the pole at $k^2=0$ (corresponding to the graviton) and \textit{no} ghost-like degree of freedom. It is worth mentioning that, given the sign difference between the spin-$2$ and spin-$0$ parts of the propagator~\eq{Propagator}, it is possible to introduce an additional healthy massive scalar in the theory, which has important applications in cosmology~\cite{Biswas:2010zk,Biswas:2011ar,Koshelev:2016xqb,Koshelev:2020xby,Koshelev:2020foq,Koshelev:2017tvv}. 

From Eq.~\eqref{ghost-free-choice} we can also obtain the corresponding relations for the form factors,
\beq
{F}_2(\Box)  = 2\frac{e^{\gamma_2(\Box)}-1}{\Box}\,,
\qquad
{F}_1(\Box)  = -\frac13 \Big[ \frac{e^{\gamma_0(\Box)}-1}{\Box} + F_2 (\Box) \Big]\,.
\eeq
Let us emphasize that the key role is played by the functions $\gamma_2$ and $\gamma_0$, which are entire; thus, the exponentials do \textit{not} introduce any unhealthy pole. Therefore, non-locality can help to resolve the ghost problem in higher-derivative gravity by means of the introduction of non-polynomial (infinite-derivative) differential operators in the action.

Finally, some simplifications can be achieved for the particular case in which $\gamma_2=\gamma_0\equiv \gamma$. In this situation the following condition holds true:
\begin{equation}
f_2(\Box) = f_0(\Box) \quad \Longrightarrow \quad {F}_1 (\Box) = -\frac{1}{2} {F}_2 (\Box) =\frac{1-e^{\gamma(\Box)}}{\Box}\,.\label{only-spin-2}
\end{equation}
Then, the non-local gravity action corresponding to the choice \eqref{only-spin-2} is given by
\begin{equation}
S = \frac{1}{2\kappa^2} \int{\rm d}^4x\sqrt{-g} \, \Big[R+G_{\mu\nu}\frac{e^{\gamma(\Box)}-1}{\Box}R^{\mu\nu}\Big]\,,
\n{only-spin-2-action}
\end{equation}
where $\,G_{\mu\nu} = R_{\mu\nu} - \tfrac12 g_{\mu\nu} R \,$ is the Einstein tensor.


\subsection{Examples of form factors}\label{sec:examples}

Here we review some explicit choices for the entire functions $\gamma_s (\Box)$ that have been intensively studied in the literature. 

\begin{enumerate}

	\item \textit{Gaussian form factor.} This is the simplest possible choice, corresponding to  the entire functions $\gamma_s(-k^2)$ that are monomials of degree one~\cite{Biswas:2005qr,Modesto:2011kw,Biswas:2011ar}:
	\begin{equation}
	\gamma_s(-k^2)= \frac{k^2}{\mu_s^2} \,,\qquad s=2,0\,,
	\label{exp-form factor}
	\end{equation}
 	where $\mu_s$ represents the physical energy scale at which non-local effects are expected to become important. 
\medskip

    \item \textit{Generic exponential form factor.} One can generalise the previous example to a generic monomial of degree $n \geqslant 1$~\cite{PU,Krasnikov,Edholm:2016hbt,hedao,Buoninfante:2018mre}: 
    \begin{equation}
    \gamma_s(-k^2)=\left( \frac{k^2}{\mu_s^2} \right) ^n\,,\qquad s=2,0\,.
    \label{exp-generic-form factor}
    \end{equation}
    The particular case $n=2$ is sometimes referred to as the Krasnikov form factor, as it was applied for the first time in gravity in Ref.~\cite{Krasnikov}.
\medskip

	\item \textit{Kuz'min and Tomboulis form factors.} This family of form factors was initially studied in Refs.~\cite{Kuzmin,Tomboulis:1997gg} and more recently in Ref.~\cite{Modesto:2011kw} and subsequent works.  A general form factor of this type can be written as
	\begin{equation}
	\gamma_s(-k^2)= \la_s \left[\Gamma\left(0,P_s(-k^2)\right)+\gamma_{\rm E}+\ln\left(P_s(-k^2)\right)\right]\,,\qquad s=2,0\,,
	\label{kuzm-form factor}
	\end{equation}
	where $\la_s$ is a positive constant, 
$\ga_{\rm E}$ denotes the Euler--Mascheroni constant, $\Gamma(0,z)$ is the incomplete gamma function and $P_s(z)$ is a real polynomial such that $\gamma_s(0)=0$. 
	The Kuz'min form factor~\cite{Kuzmin} consists in the choice $P_s(z) = -z /\mu_s^2$ with an arbitrary $\la_s \in \mathbb{N}$, while the Tomboulis form factor~\cite{Tomboulis:1997gg} is characterized by $\la_s = 1/2$ and $P_s(z)$ of arbitrary (but even) degree $N_s \geqslant 2$.
	Despite their more complicated expressions (in comparison to the previous ones), they are useful to formulate super-renormalizable non-local theories of gravity because, for large momentum, they behave like a polynomial. In fact~\eq{kuzm-form factor} yields
\beq
e^{\gamma_s(-k^2)} \underset{k^2 \to \infty}{\sim} \, k^{2 \la_s N_s},
\eeq
where $N_s$ is the degree of the polynomial $P_s(z)$.
\end{enumerate}


\section{Stationary linearised solutions}
\label{stationary sol}

Most of the solutions in non-local gravity were obtained in the linear approximation. In this section we describe two types of such solutions, namely, the Newtonian-limit solution and the weak-field slowly rotating stationary solution. To this end, let us consider a generic metric with line element 
\begin{equation}
{\rm d}s^2=-(1+2\ph){\rm d}t^2+2\vec{h}\cdot {\rm d}\vec{r} \, {\rm d}t +(1-2\psi)({\rm d}x^2+{\rm d}y^2+{\rm d}z^2)\,,\label{isotr-metric}
\end{equation}
where $\vec{r}=(x,y,z)$ is the radius vector, and $\ph = -\kappa h_{00}/2,$ $\psi  = - \kappa h_{ii}/2$ and $h_i = \kappa h_{0i}$ are the potentials sourced by $T_{\mu\nu}.$ 

In the non-relativistic limit, the source is pressureless, $T\equiv \eta^{\mu\nu}T_{\mu\nu}\simeq -T_{00}$, and its only non-vanishing components are $T_{00}$ and, possibly, $T_{0i}$. The latter is zero for static point-like sources (so that in this case $h_i \equiv 0$) and it is non-zero for a rotating dust. 

The metric potentials $\ph$ and $\psi$ can be obtained from the 00-component and the trace of the linearised field equations~\eqref{lin-field-eq}, which are equivalent to 
the set of coupled differential equations:\footnote{In the case of $T_{0i}\neq 0$, the de Donder gauge $\partial_\mu h^{\mu\nu}=0$ is assumed, or a suitable higher-order generalisation compatible with the metric~\eqref{isotr-metric} (see, \textit{e.g.},~\cite{Accioly:2016qeb,Teyssandier89,Buoninfante:2018xif,BuoGia}).}
\begin{subequations} \label{field-eq-pot}
\begin{align} 
\left[ a(\De)-c(\De)\right] \De \ph + 2 c(\De) \De \psi =  & \,\, \kappa^2 \,T_{00}\,, 
\label{field-eq-pot1}
\\[2mm]
\left[ 3c(\De) - a(\De)\right]  \left[ 2 \De \ph - \De\psi\right]  =  & \,\, \kappa^2 \,T_{00}\,, 
\label{field-eq-pot2}
\\[2mm]
a(\De)\,\De h_{i} = & \, - 2\kappa^2\,T_{0i}\,, 
\label{field-eq-pot3}
\end{align}
\end{subequations}
where the functions $a$ and $c$ now depend on the spatial Laplacian $\De$, as $\Box\simeq \De$ for stationary fields.
Notice that for $a=c=1$ ({\it i.e.}, ${F}_1={F}_2=0$) the system~\eqref{field-eq-pot} reduces to standard Poisson equations, consistently recovering the Einstein's GR result.

The first two equations in~\eqref{field-eq-pot} can be decoupled by using the auxiliary potentials~\cite{BreTib1}
\begin{equation}
\chi_2 \equiv \frac{\ph+\psi}{2} \qquad \textrm{and} \qquad    \chi_0\equiv 2\psi - \ph \,.
\end{equation}
In fact, the field equations~\eqref{field-eq-pot} are equivalent to
\beq
f_s(\De) \De \chi_s   & = &  \frac{\kappa^2}{2} T_{00}\,, \label{chi_s diff} 
\\[2mm]
f_2(\De) \De h_i  & = & -2 \kappa^2 T_{0i}\,,\label{h_i diff}
\eeq
where $s=0,2$ and the functions $f_s$ were defined in Eq.~\eq{f2-f0}. Once the above equations are solved for $\chi_{0,2}$, the original potentials are obtained through the relations
\beq
\label{pch}
\ph =  \frac43 \chi_2 - \frac13 \chi_0
, \qquad  \qquad
\psi = \frac23 \chi_2 + \frac13 \chi_0
.  
\eeq

The functions $\chi_{s}$ are called spin-$s$ potentials because they only depend on the spin-$s$ part of the field. In fact, each $\chi_{s}$ only depends on $f_s(\De)$, which in its turn is related to the gauge-invariant spin-$s$ part of the propagator~\eqref{Propagator}.
Also, notice that for the particular case described in~\eq{only-spin-2} we have $a=c=f_2=f_0$ and all the Newtonian potentials are equal, namely,
\beq
\label{pp=cc}
\ph = \psi = \chi_2 = \chi_0.
\eeq 

In what concerns the off-diagonal components $h_{i}$ of the metric, as Eq.~\eqref{h_i diff} only depends on $f_2$, they are  
not affected by  ${F}_1$ in the linear regime. This is due to the fact that the form factor ${F}_1$ modifies only the scalar part of the propagator, which couples to the trace of the energy-momentum tensor. Thus, since the components $T_{0i}$ do not enter in the trace $g^{\mu\nu}T_{\mu\nu}$, they cannot act as sources for $h_{i}$ at this order in perturbation.

In what follows we show some general results on the solutions for the Newtonian potentials $\chi_{s}$, and we postpone the consideration of the off-diagonal components $h_{i}$ to Sec.~\ref{SecRing}.


\subsection{Solution for the potentials in terms of effective delta sources}
\label{SecEffS}

It is possible to solve Eq.~\eq{chi_s diff} directly using the Fourier transform method, or the Laplace transform (see Sec.~\ref{eat-kernel} below), but in some cases it is instructive to consider an intermediate step, which we describe here (see also, \textit{e.g.},~\cite{BreTib2,Nos6der}). 

An alternative interpretation of Eq.~\eq{chi_s diff} follows from the inversion of the operator $f_s (\De)$, so that one obtains a standard Poisson equation with a modified source,
\beq
\n{ED-Source}
\De \chi_s = 4 \pi G \rho_s .
\eeq
Accordingly, the \emph{effective source} $\rho_s$ satisfies 
\beq
\label{InvFs}
\rho = f_s(\De) \, \rho_s,
\eeq
where $\rho = T_{00}$ is the mass density.

For a point-like source with mass $m$, $\rho(\vec{r}) = m \de^{(3)} (\vec{r})$ and the \emph{effective delta source} can be obtained via the Fourier transform method applied to Eq.~\eq{InvFs}. The result is
\beq
\n{lap-eff}
\rho_s (r) = \frac{m}{2 \pi^2} \int_0^\infty \rd k \, \frac{k \sin(kr)}{r f_s(-k^2)}
,
\eeq
where $r = |\vec{r}|$. 

The solution of Eq.~\eq{ED-Source} for the potential $\chi_s$ with the effective delta source~\eq{lap-eff} then reads 
\beq
\label{chiIntg}
\chi_s (r) = G \int_\infty^r \rd x \, \frac{m_s(x)}{x^2}
,
\eeq
where $m_s(r)$ is the \emph{effective mass function},
\beq
\label{massfunction}
m_s(r) = 4\pi \int_0^r \rd x \, x^2 \rho_s (x) 
.
\eeq

The formulation of the Newtonian limit of higher-derivative gravity models in terms of the effective delta source~\eq{lap-eff} has been fruitfully applied to local and non-local models (see, \textit{e.g.},~\cite{Tseytlin95,BreTib2,Nos6der}). The main advantage of the method is the possibility of deriving general results about the regularity of the potentials that depend only on the behaviour of the form factor in the UV regime, as we discuss in Sec.~\ref{SecPon} below.


\subsection{Heat kernel solution for the Newtonian potentials}
\label{eat-kernel}

In the previous subsection, we showed how the solution for the potential $\chi_s$ can be obtained in the effective source formalism. Here, we present another integral solution for the potential, using the heat kernel approach based on the Laplace transform, as carried out in~\cite{Frolov:Poly,Frolov:Exp}. This solution is particularly useful in the context of~Sec.~\ref{mini-bh}, when dealing with the dynamic process of mini black hole formation. 

Introducing the Green's function for~\eq{chi_s diff} via
\beq
\n{Green-function}
f_s (\De) \De \,  G_s ( \vec{x},\vec{x}') \,=\, \de^{(3)} (\vec{x}- \vec{x}') 
\,, 
\eeq
the integral solution for the potential is given by
\beq
\label{cGF}
\chi_{s} ( \vec{x}) \,=\, 4 \pi G 
\int \rd ^3 x' \, G_{s} ( \vec{x},\vec{x}') \, \rho (\vec{x}')
\,.
\eeq
Let us now assume that the inverse of the differential operator in~\eq{Green-function} can be written as the Laplace transform of a function $h_{s}(s')$, that is,
\beq 
\label{H-1}
-  \frac{1}{\xi f_{s}(-\xi)} \,=\, \int_0^\infty \rd s' \, h_{s}(s')\, e^{-s' \xi} 
\,.
\eeq
Then, the $x$-representation of the Green's function reduces to
\beq 
\n{Gxx}
G_{s} ( \vec{x},\vec{x}') \,=\, \int_0^\infty \rd s' \, h_{s} (s') \, 
\langle \vec{x} \,| \, e^{s' \, \De} \,|\, \vec{x}' \rangle
\,,
\eeq
where 
\beq
\langle \vec{x} \,| \, e^{s \, \De} \,|\, \vec{x}' \rangle
\,=\, K(|\vec{x}-\vec{x}'|;s) 
\,=\, \frac{ 
e^{ - \frac{|\vec{x}-\vec{x}'|^2}{4s} } 
}
{(4\pi s)^{ \frac32}}
\, 
\eeq
is the heat kernel of the Laplacian. 

For a point-like source $\rho (\vec{x}) = m \de^{(3)} (\vec{x})$ the formula \eq{cGF} simplifies to
\beq
\label{cHK}
\chi_{s} (r) \,=\, 4 \pi G m 
\int_0^\infty \rd s'\, h_{s}(s')\, K(r;s'), \qquad s= 0,2
\,,
\eeq
where $r = |\vec{x}|$. Explicit calculations using this method can be found in~\cite{Frolov:Poly,Frolov:Exp,BreTib1,BFZ-pbranes,hedao} and in the Sec.~\ref{egnonlocal} below.


\subsection{Resolution of point-like singularities}
\label{SecPon}

The resolution of the Newtonian point-like singularity in non-local gravity models has been studied by several authors~\cite{Tseytlin95,Modesto:2011kw,Modesto:2012ys,Jens,Edholm:2016hbt,hedao,Biswas:2011ar,Nos6der,Nos4der,BreTib2,Buoninfante:2018xiw,Buoninfante:2018rlq,Buoninfante:2018stt,Ercan18,BFZ-pbranes,Buoninfante:2020ctr,Buoninfante:2018lnh,Abel:2019zou,delaCruz-Dombriz:2018aal,Kolar:2020bpo}. In most of the publications, a specific form factor was chosen and then the potentials (and, sometimes, also some curvature invariants) were evaluated, either analytically or numerically. Since there is an infinite amount of non-local form factors and of curvature invariants, it is impossible to proceed the investigation one by one. This issue was solved in~\cite{Nos6der}, by using the effective source formalism (see Sec~\ref{SecEffS}) and  presenting necessary and sufficient conditions that a form factor must satisfy for the Newtonian limit to be regular --- as defined in terms of gravitational potential, curvature or curvature-derivative invariants.\footnote{We use the term \textit{curvature-derivative invariant} to a generic scalar which is polynomial on curvature tensors and their covariant derivatives.}

Regarding the regularity of the potentials, if the function $f_s(-k^2)$ grows at least as fast as fast as $k^{2}$ for large $k$, the smearing of the original delta source through~\eq{lap-eff} is enough to yield a potential $\chi_s(r)$ that is finite at $r=0$~\cite{BreTib2}. This happens even in the case of $f_s(-k^2) \sim k^2$, for which the effective delta source $\rho_s(r)$ is unbounded (see~\cite{Nos4der,Nos6der}). In most cases, however, the effective delta source is also bounded; in fact, $\rho_s(r)$ is finite if $f_s(-k^2) \sim k^4$ or faster~\cite{BreTib2}. If both $\chi_0$ and $\chi_2$ are regular, so are the potentials $\ph$ and $\psi$ [see Eq.~\eq{pch}].  In this spirit, one might say that the improved UV behaviour of the propagator~\eq{Propagator} regularizes the Newtonian singularity of the potentials. 
We stress that the resolution of this divergence is \emph{not} owing to the non-locality of the theory; on the contrary, it was first noticed in the local fourth-derivative gravity~\cite{Stelle77} and later in other more general local higher-derivative models~\cite{Newton-MNS,Giacchini:2016xns,BreTib1}. 

It is possible to achieve an increasing degree of regularity by including more derivatives and non-localities in the action. This can be better formulated in terms of the following definition.
\begin{definition}[\normalfont Order of regularity of a potential~\cite{Nos6der}] 
If there exists a $p\in\mathbb{N}$ such that $\chi_s(r)$ is at least $2p$-times continuously differentiable, $\chi_s^{(2p)}(r)$ is bounded and $\chi_s^{(2n+1)}(0) = 0$ for all $n\in\lbrace 0,\ldots,p-1\rbrace$, then $p$ is called the \emph{order of regularity} of $\chi_s(r)$ (equivalently, it is said that $\chi_s(r)$ is \emph{$p$-regular}).
\end{definition}

With this definition, as the next theorem shows, the considerations about the usual regularity of the potential can be extended to the regularity of the curvature invariants associated with the static (\textit{i.e.}, with $h_i \equiv 0$) metric~\eq{isotr-metric}.
\begin{theorem}[\normalfont Regularity of linearised curvature-derivative invariants~\cite{Nos6der}]
\label{Theorem1}
Let the form factors be such that $f_s(-k^2)$ asymptotically grows at least as fast as $k^{4+2N_s}$ for an integer $N_s\geqslant 0$ and let $N \equiv \min \lbrace N_0, N_2 \rbrace$. Under these conditions, the potential $\chi_s(r)$ is at least $(N_s+1)$-regular and all the linearised curvature-derivative invariants\footnote{That is, curvature-derivative invariants calculated with the metric~\eq{isotr-metric}, with $h_i \equiv 0$, to the leading order in the metric perturbation.} with at most $2N$ covariant derivatives are bounded.
\end{theorem}
\begin{corollary}
If $f_{2}(-k^2)$ and $f_{0}(-k^2)$ grow faster than any polynomial, then the potentials $\chi_{0,2}(r)$ are even analytic functions of $r$ and all the curvature invariants with an arbitrary number of covariant derivatives are regular.
\end{corollary}
The proof of the theorem can be found in~\cite{Nos6der}, where the regularity properties of the potential $\chi_s$ were deduced from some basic features of $f_s(-k^2)$ translated into the effective delta source~\eq{lap-eff}.

If the potentials $\chi_{0,2}$ are only 0-regular (like in the case of the Kuz'min or Tomboulis form factors~\eq{kuzm-form factor} that asymptotically tend to $k^2$), then the metric still has curvature singularities. On the other hand, if the potentials are only 1-regular, all the curvature invariants without covariant derivatives (such as $R$ and $R_{\mu\nu\al\be}R^{\mu\nu\al\be}$) are regular, but there are scalars with two covariant derivatives that diverge, such as $\Box R$ --- this is the case, \textit{e.g.}, of the form factors~\eq{kuzm-form factor} that tend to $k^4$. 

Hence, besides characterizing the models that have a regular Newtonian limit, the theorem also shows that the regularity of curvature invariants at $r=0$ depends on the absence of odd powers of $r$ in the Taylor expansion of the potentials $\chi_s(r)$. The higher the first odd-order term is, the larger the set of regular scalars is. 
A simple example is provided by the family of scalars $\Box^n R$ (for an arbitrary $n$), which in the Newtonian limit reads
\beq
\label{LapN_R}
(\Box^n R)_\text{lin} 
=
2 \De^{n+1} \chi_0
=  2 \left[ \chi_0^{(2n+2)} + \frac{2(n+1) }{r} \chi_0 ^{(2n+1)}  \right] 
.
\eeq 
For the above scalar to be regular, it suffices that the potential $\chi_0(r)$ be $(n+1)$-regular --- in other words, that $\chi_0(r)$ is $(2n+2)$-times continuously differentiable and the limit $\lim_{r\to 0} \chi_0 ^{(2n+1)}(r)/r$ exists~\cite{Nos6der}. If the former condition is verified, but the latter is not, then $(\Box^n R)_\text{lin}$ has a singularity at $r=0$. Only if the form factor grows faster than any polynomial [such as~\eq{exp-generic-form factor}], all the local curvature-derivative invariants are regular. 

Finally, we remark that the Theorem~\ref{Theorem1} does not distinguish between local and non-local higher-derivative models, as it only depends on the function $f_s(-k^2)$. Therefore, the regularization of the Newtonian limit can be achieved in both types of theories, and it is not a consequence of the absence/presence of ghosts or non-localities~\cite{BreTib2}. 
Nevertheless, the $\infty$-order regularity can only be achieved in non-local models, but not in all of them. Namely, only in the ones with form factors that grow faster than any polynomial [\textit{e.g.},~\eq{exp-generic-form factor}]; in this case, the effective delta source~\eq{lap-eff} and the associated Newtonian potential $\chi_s(r)$ are even analytic functions of $r$~\cite{Nos6der}.


\subsubsection{Example 1: Gaussian form factor}
\label{SecExExp}

\begin{figure}[t]
\begin{center}
\includegraphics[scale=.6]{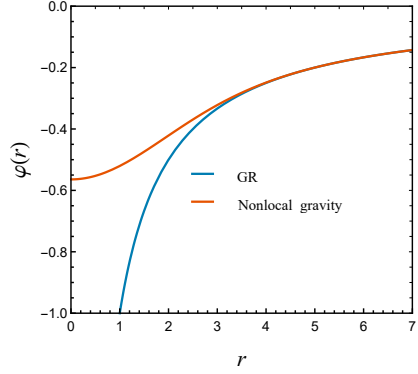}
\caption{Behaviour of the metric potential $\ph(r)$ for the non-local gravity model in Eq.~\eqref{static-neutral-solution} (red line) and GR (blue line). For convenience, we have set $\mu= G = m =1$.
}
\label{fig1}     
\end{center}
\end{figure}

As an explicit example, let us consider the form factor~\eq{exp-form factor} with $f_0(-k^2) = f_2(-k^2) = \exp(k^2/\mu^2)$.
The Eq.~\eq{pp=cc} is valid in this case, and the potentials can be obtained through the methods described in the previous sections; the solution reads~\cite{Tseytlin95,Siegel:2003vt,Biswas:2005qr,Modesto:2011kw,Biswas:2011ar}
\beq
\ph(r) = -\frac{Gm}{r}{\rm erf}\left(\frac{\mu r}{2}\right)\,,
\label{static-neutral-solution}
\eeq
where $\text{erf}(x)$ 
is the error function. 
As expected from the above considerations, the gravitational potential in Eq.~\eqref{static-neutral-solution} is non-singular at $r=0$ and tends to the finite constant value $\ph(0)=Gm\mu/\sqrt{\pi}$, whereas in the large-distance limit ($r\mu\gg 1$)  the $1/r$ behaviour of the Newtonian potential is recovered (see Fig.~\ref{fig1}). 

Moreover, since the functions $f_{0,2}(-k^2)$ grow faster than any polynomial, 
all the linearised curvature invariants (even those constructed with covariant derivatives of the curvatures) are bounded. For instance, we can compute the linearised Kretschmann invariant~\cite{Buoninfante:2018xiw},
\begin{equation}
\hspace{-0.1cm}
\,\, (R_{\mu\nu\al\be}^2)_{\rm lin} = \frac{G^2m^2}{3\pi r^6} e^{-\frac{\mu^2r^2}{2}} \Big\lbrace 5\mu^6r^6+ 4\Big[6\mu r+\mu^3r^3-6\sqrt{\pi}e^{\frac{\mu^2r^2}{4}}{\rm erf}\Big(\frac{\mu r}{2}\Big)\Big]^2\Big\rbrace\,,
\end{equation}
and explicitly verify that it is non-singular. Indeed,
\begin{equation}
\lim\limits_{r\rightarrow 0} (R_{\mu\nu\al\be}^2)_{\rm lin} = \frac{5G^2m^2\mu^6}{3\pi}\,,
\end{equation}
while, in the large-distance limit it consistently recovers the GR expression 
\beq
(R_{\mu\nu\al\be}^2)_{\rm lin}\sim \frac{48G^2m^2}{r^6}.
\eeq

From a physical point of view, the regularization that we have just showed can be interpreted as if the non-local form factor smears out the point-like source at $r=0$ on a spatial region of size $\sim 1/\mu$. 
In addition, in~\cite{Buoninfante:2018rlq,Buoninfante:2018xiw} it was shown that the above regular metric becomes conformally flat in the limit $r\rightarrow 0$, as the Weyl tensor vanishes at the origin; more generally, this happens provided that the potential $\chi_2(r)$ is at least 1-regular~\cite{BreTib1}.

Similar qualitative results hold for the form factors in Eq.~\eq{exp-generic-form factor}, which also grow faster than any polynomial. See, \textit{e.g.},~\cite{Edholm:2016hbt,hedao,Jens,BFZ-pbranes} for other explicit examples of this type.


\begin{figure}[t]
	\begin{center}
		\includegraphics[scale=.42]{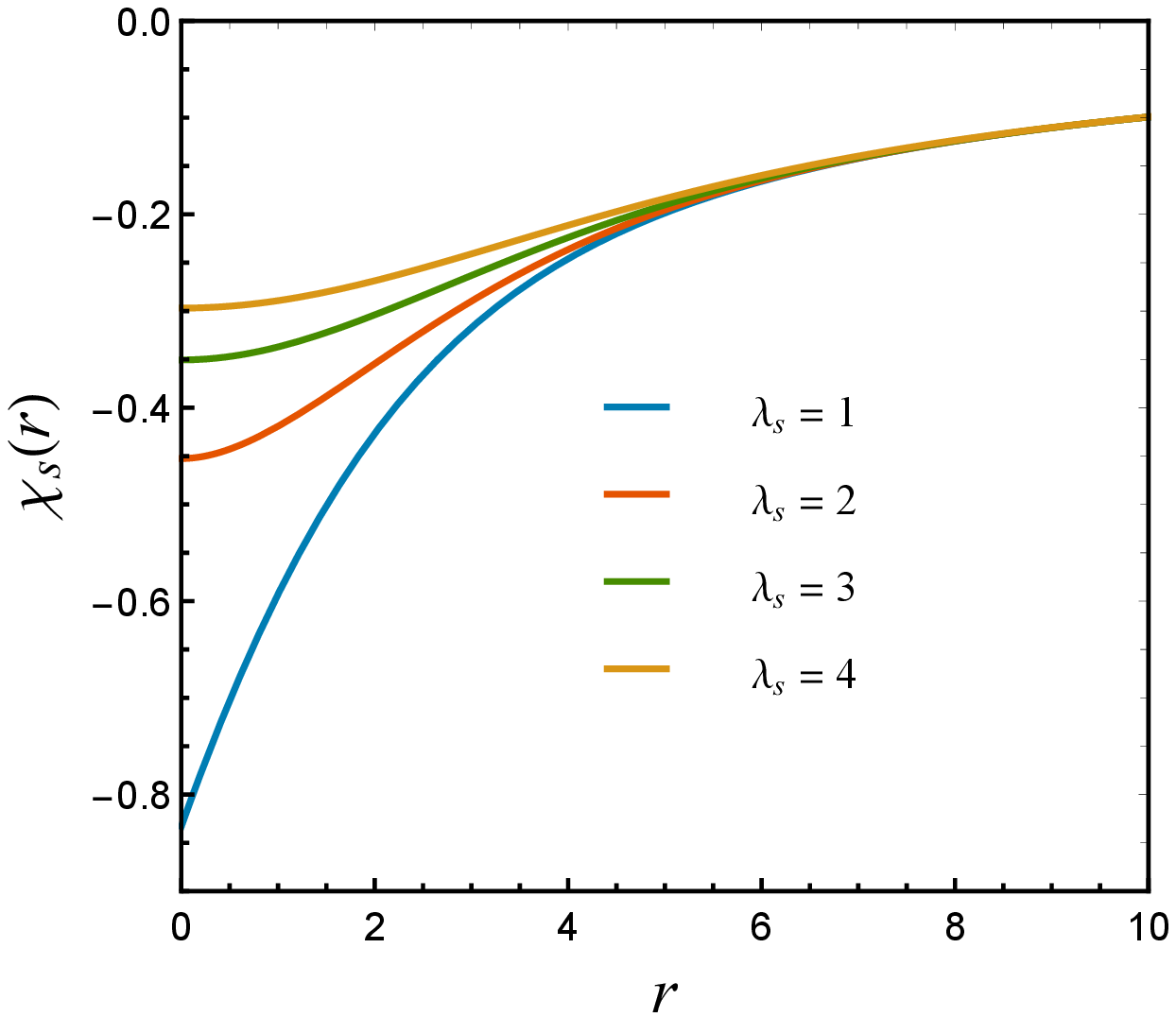}
		\hspace{0.5cm}
		\includegraphics[scale=.42]{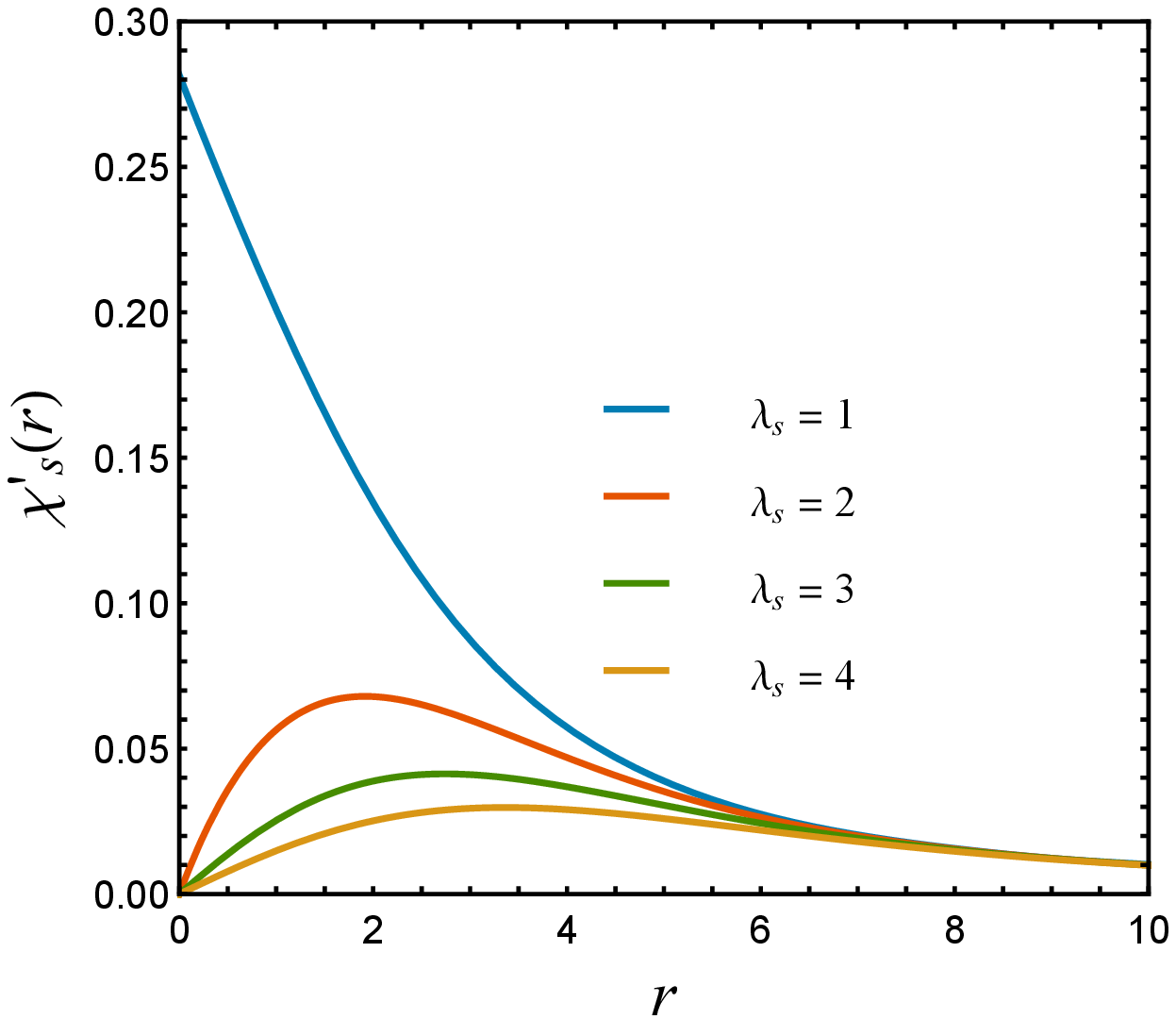}
		\\
		\vspace{0.4cm}
		\includegraphics[scale=.42]{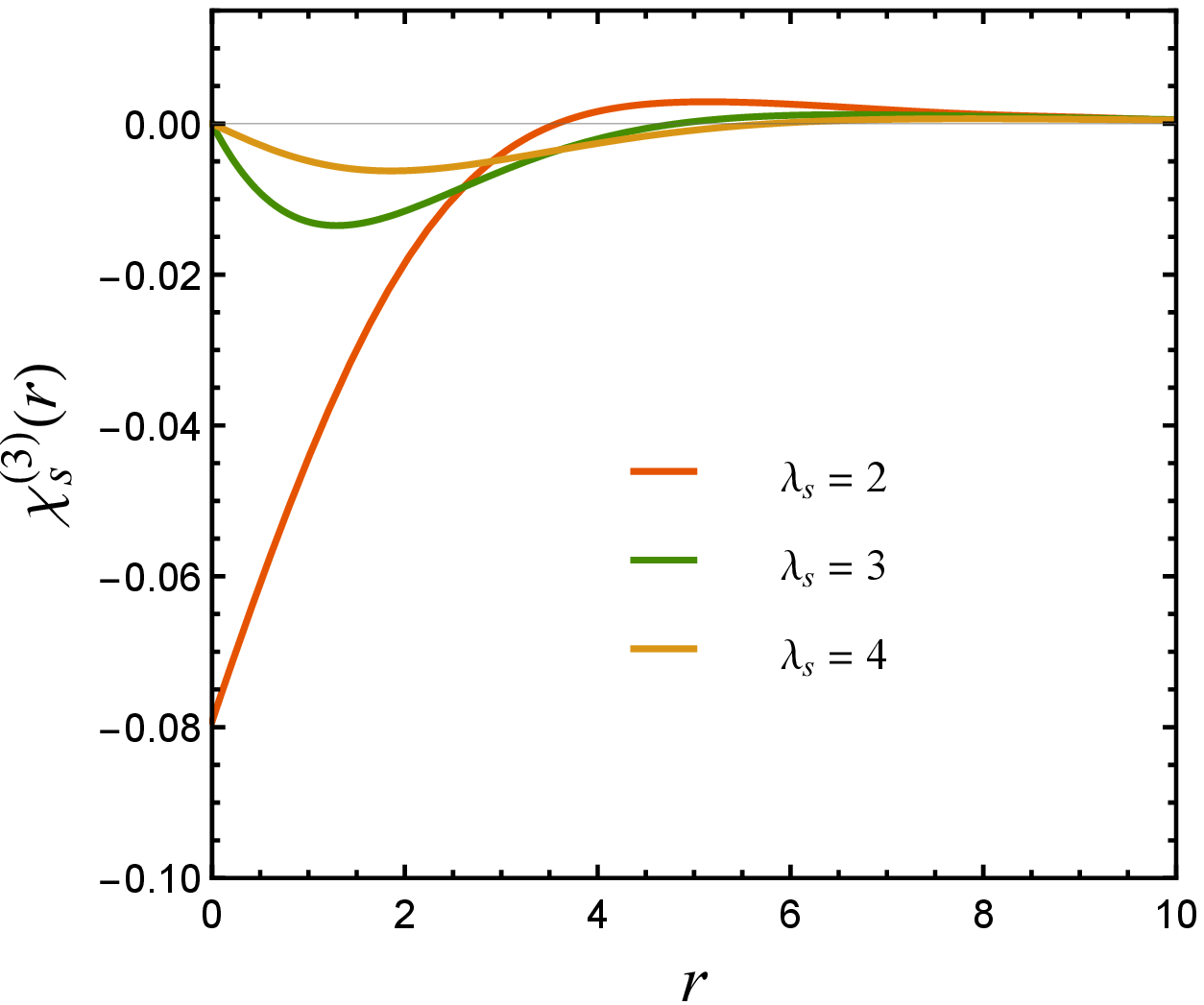}
		\hspace{0.4cm}
		\includegraphics[scale=.42]{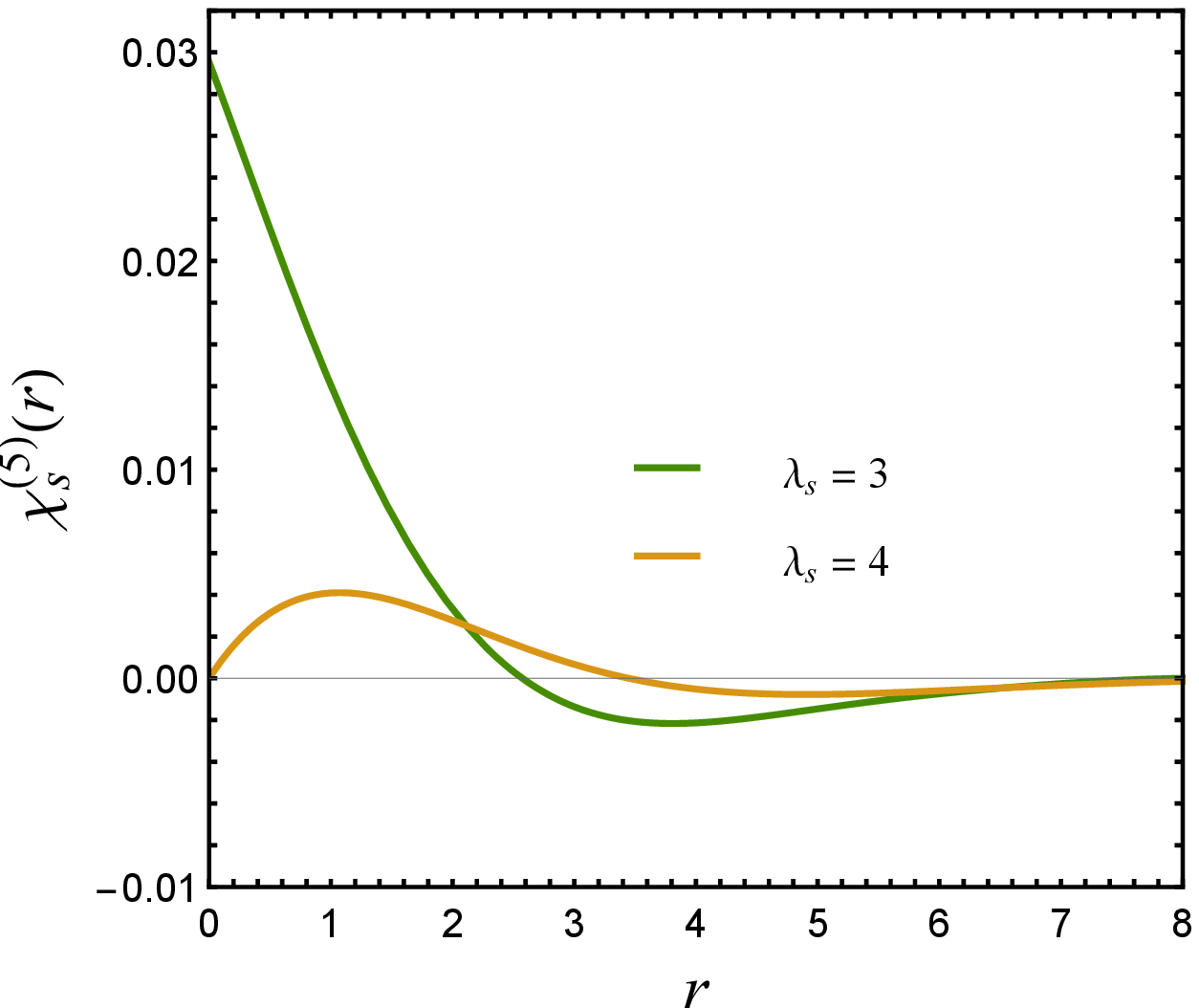}
		\hspace{0.1cm}
		\caption{Numerical evaluation of $\chi_s(r)$ and its first odd-order derivatives for the Kuz'min form factor with $\la_s\in\lbrace 1,2,3,4\rbrace$ in~\eqref{HKuz}; we have set $m = G = \mu_s = 1$.}
		\label{Graph1}
	\end{center}
\end{figure}

\subsubsection{Example 2: Kuz'min form factor}

Since the previous example dealt with a form factor that grows faster than any polynomial, here we discuss the case (presented in~\cite{Nos6der}) of a form factor that behaves like a polynomial in UV regime. To this end, let us consider the Kuz'min form factor~\eq{kuzm-form factor}~\cite{Kuzmin}, for which
\beq
\label{HKuz}
\ga_s(-k^2) = \lambda_s \left[ \ga_{\rm E} + \Ga(0,k^2/\mu_s^2) + \ln (k^2/\mu_s^2) \right] ,
\eeq 
where $\mu_s$ is a mass parameter and $\lambda_s \in \mathbb{N}$. For large momentum it satisfies 
\beq
\label{HKuzLim}
f_s(-k^2)  \, \underset{k^2 \to \infty}{\approx} \, e^{\lambda_s \ga_{\rm E}} \,  \left( \frac{k^2}{\mu_s^2}\right) ^{ \lambda_s}  \, ,
\eeq
thus $f_s(-k^2) \sim k^{2\la_s}$ for $k$ large enough. From the aforementioned results relating the UV behaviour of the form factor and the order of regularity of the Newtonian potentials~\cite{Nos6der} 
it follows that, if $\lambda_s = 1$, the effective delta source is still singular and the 
associated potential is only 0-regular. On the other hand, the effective delta source is bounded for $\la_s \geqslant 2$, and the potential $\chi_s(r)$ is $(\la_s-1)$-regular.

These features can be viewed in Fig.~\ref{Graph1}, which displays the numerical evaluation of $\chi_s(r)$, $\chi_s^\prime(r)$, $\chi_s^{(3)}(r)$ and $\chi_s^{(5)}(r)$ for the parameter $\la_s \in \lbrace 1, 2, 3, 4 \rbrace$. Notice that the potential is finite for $\la_s=1$, but its first derivative does not vanish at $r=0$, indicating the singularity of the source (and of the curvature invariants).
On the other hand, for $\lambda_s \geqslant 2$ the corresponding potential is at least $(\la_s-1)$-regular, as discussed above. More precisely, we see that $\chi_s^{(2\la_s-1)}(0)\neq 0$, showing that the potential is not $\la_s$-regular.

Similar results also hold for the form factors considered by Tomboulis~\cite{Tomboulis:1997gg} and Modesto~\cite{Modesto:2011kw}, which behave like polynomials in the UV. See~\cite{BreTib2} for another example.


\subsection{Resolution of ring-like singularities}
\label{SecRing}

The previous discussion was focused on static solutions; here we return to the more general metric in the form~\eqref{isotr-metric} with non-trivial $h_i$ and show that non-locality is also able to regularize rotating ring-like singularities. We follow the presentation of~\cite{Buoninfante:2018xif} and only consider the particular choice of form factor~\eq{exp-form factor} such that $f_0(-k^2) = f_2(-k^2) = \exp(k^2/\mu^2)$, like in the example of Sec.~\ref{SecExExp}.

In GR, the Kerr metric is plagued by a ring singularity that in Boyer-Lindquist coordinates is described by the equation $r^2+a^2{\rm cos}^2\vartheta=0$ or, in Cartesian coordinates, by $z=0$ and $x^2+y^2=a^2,$ where $a$ can be thought of as the radius of the ring.
We can mimic such a ring distribution by modelling the energy-momentum tensor as a Dirac delta distributed on a ring of radius $a$ rotating with constant angular velocity $\omega$~\cite{Buoninfante:2018xif}:
\begin{equation}
T_{00}=m\delta(z)\frac{\delta^{(2)}(x^2+y^2-a^2)}{\pi}, \qquad T_{0i}=T_{00} v_i, \label{ring}
\end{equation}
where $v_i$ is the tangential velocity and its magnitude satisfies the relation $v=\omega \,a.$ By taking the $z$-axis as the direction of the angular-velocity vector, we can write
$v_x=-y \,\omega,$ $v_y=x \,\omega$ and $v_z=0.$ Note that, in the limit $a\rightarrow 0$ we consistently recover the expression of the point-like source because $v_i\rightarrow 0$ and $\delta^{(2)}(x^2+y^2-a^2)\rightarrow \delta^{(2)}(x^2+y^2)=\pi \delta(x)\delta(y).$

In this configuration, the off-diagonal elements of the metric~\eqref{isotr-metric} are non-vanishing, and the set of non-local differential equations~\eqref{field-eq-pot} reduces to
\begin{equation}
\begin{array}{rl}
\displaystyle e^{-\De/\mu^2}\De \ph (\vec{r}) = \displaystyle e^{-\De/\mu^2}\De \psi(\vec{r}) = & 4Gm \,\delta(z)\delta^{(2)}(x^2+y^2-a^2),
\\[2.5mm]
\displaystyle e^{-\De/\mu^2}\De h_{0x}(\vec{r})=& \displaystyle-16Gm\,\omega \,y \delta(z)\delta^{(2)}(x^2+y^2-a^2),
\\[2.5mm]
\displaystyle e^{-\De/\mu^2}\De h_{0y}(\vec{r})=& \displaystyle 16Gm\,\omega\, x \delta(z)\delta^{(2)}(x^2+y^2-a^2).
\end{array}\label{diff-eq}
\end{equation}
We can use the Fourier transform method to solve the modified Poisson equations. It turns out to be useful to work in cylindrical coordinates, {\it i.e.}, $x=\rho \cos \varphi,$ $y=\rho \sin \varphi,$ $z=z.$ The Fourier transform of the energy-momentum tensor components read~\cite{Buoninfante:2018xif}
\beq
\mathcal{T}[\delta(z)\delta^{(2)}(x^2+y^2-a^2)] & = &   \pi \, I_0\left(ia\sqrt{k_x^2+k_y^2}\right),
\label{fourier00}
\\
\mathcal{T}[x\delta(z)\delta^{(2)}(x^2+y^2-a^2)] & = & \pi   a \frac{k_x}{\sqrt{k_x^2+k_y^2}} I_1\left(ia\sqrt{k_x^2+k_y^2}\right),
\\
\mathcal{T}[y\delta(z)\delta(x^2+y^2-a^2)] & = &\pi a \frac{k_y}{\sqrt{k_x^2+k_y^2}} I_1\left(ia\sqrt{k_x^2+k_y^2}\right),
\eeq
where $\mathcal{T}[\cdots]$ stands for the Fourier transform operation, $I_0$ and $I_1$ are modified Bessel functions of the first kind. 

For simplicity, let us analyse the behaviour of the solution in the $xy$-plane ({\it i.e.} $z=0$), where we expect the singularity to appear in GR, and work with the cylindrical radial coordinate $\rho=\sqrt{x^2+y^2}$. Therefore, by transforming back to coordinate space, the metric potentials are given by~\cite{Buoninfante:2018xif}
\begin{equation}
\ph(\rho)=\psi(\rho)= -Gm \int_{0}^{\infty}\rd \zeta \, I_0\left(ia\zeta\right)I_0\left(i\zeta \rho \right){\rm erfc}\left(\frac{\zeta}{\mu}\right)\,,
\label{grav-potIDGnum}
\end{equation}
\begin{equation}
h_{0x}(x,y)= 4Gm\,\omega \,a \frac{y}{\rho}H(\rho), \qquad
h_{0y}(x,y)= -4Gm\,\omega \,a \frac{x}{\rho}H(\rho),\label{h0y}
\end{equation}
where ${\rm erfc}(z)$ is the complementary error function and 
\begin{equation}
H(\rho) \equiv \int_{0}^{\infty} \rd\zeta \, I_1(ia\zeta)I_1(i\zeta \rho) {\rm erfc}\left(\frac{\zeta}{\mu}\right)\,.
\label{H-IDG}
\end{equation}

\begin{figure}[t]
\begin{center}
	{	\includegraphics[scale=0.55]{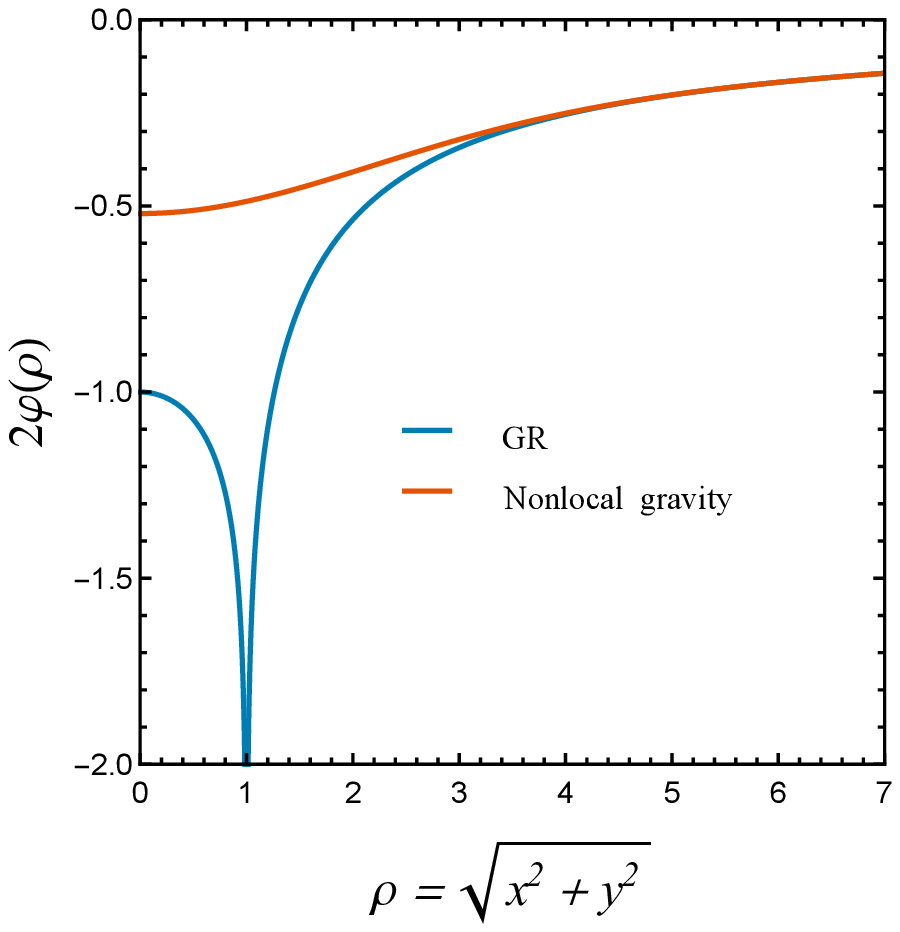}
		\qquad\quad
		\includegraphics[scale=0.55]{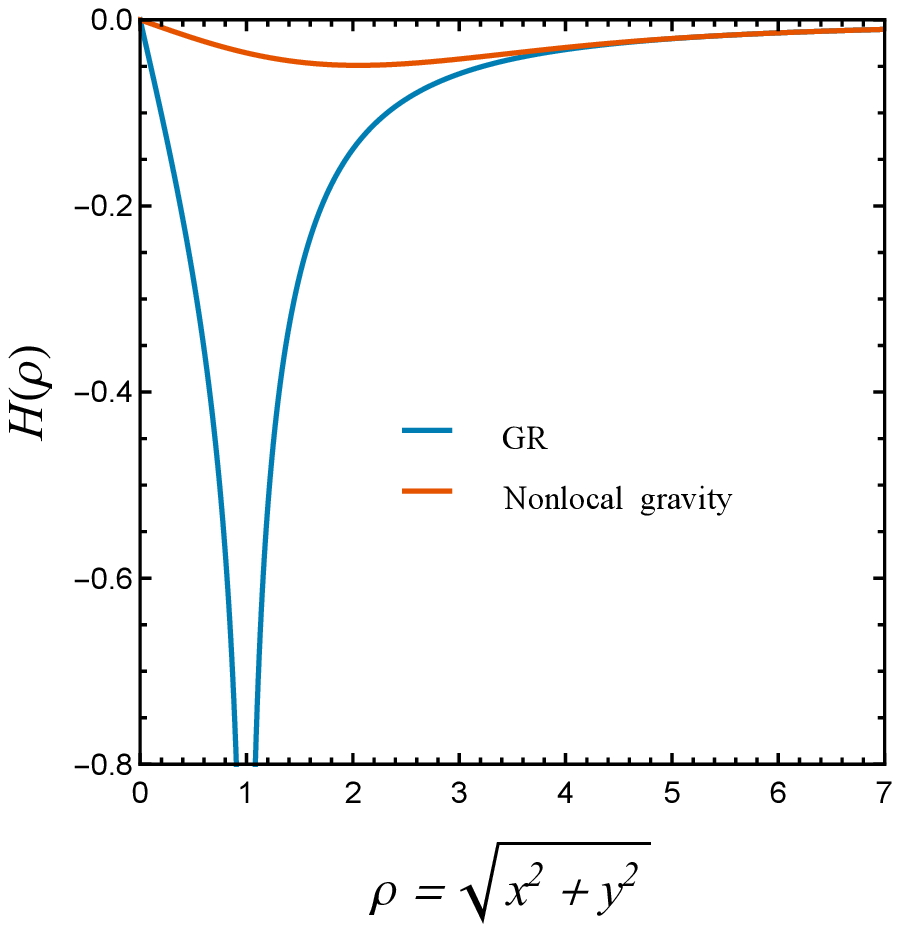}
	}
	\caption{Behaviour of the components $-h_{00}=2\ph$ and $h_{0i}\sim H$ as functions of the cylindrical radius $\rho,$ for the non-local gravity model (red line) and GR (blue line). For convenience, we have set $\mu=1,$ $2Gm=1$ and $a=1.$}
	\label{fig2}
\end{center}
\end{figure}

The above integrals cannot be solved analytically, but the integrations can be performed numerically and the solutions are plotted in Fig~\ref{fig2}. It shows that in GR the metric potentials are singular at $\rho=a,$ as expected, whereas in the non-local gravity model under investigation the would-be ring singularity is regularized. Also in this case, one can verify that all curvature invariants are non-singular in the entire spacetime and that the Weyl tensor vanishes at $r=0$~\cite{Buoninfante:2018xif}.

Let us remark that the linearised regime holds as long as the conditions $2\vert \ph \vert<1$ and $\vert h_{0i} \vert <1$ are satisfied. In particular, they are valid for any value of $r$ if, neglecting constant factors of order one, 
$Gm \mu < 1$ and $Gm\mu^2\omega a^2 < 1,$ respectively~\cite{Buoninfante:2018xif}. From the last inequality it is evident that the angular velocity $\omega$ cannot be too large otherwise the linear approximation would break down. Finally, the metric outside the source was obtained using the multipole expansion also in~\cite{Buoninfante:2018xif}, and the extension of this procedure for more general form factors was carried out in~\cite{BuoGia}.


\section{Mini black hole formation by the collapse of null shells}
\label{mini-bh}

Up to this point, our considerations were restricted to linearised stationary configurations without horizons. In this section, we discuss the collapse of small-mass spherical null shells. By small mass we mean that we continue to work with the linearised equations for the gravitational field. The interest in this scenario is the possibility of the emergence of mini black holes. 

The gravitational field of a null shell can be built in the linearised limit using the {\it superposition principle}. By boosting the weak-field potential $\chi_s$ of a static point-like mass, the gravitational field of an ultrarelativistic particle is obtained, and taking a spherical superposition of such solutions, the linearised metric of a collapsing null shell is constructed. The formalism we follow was introduced in~\cite{Frolov:Exp}, where it was applied to non-local gravity with form factor $f_s(-k^2) = \exp (k^2/ \mu_s^2 )$. It was later generalised to an arbitrary form factor $f_s (-k^2)$ in~\cite{Frolov:Poly} and applied to the case of local higher-derivative polynomial models in~\cite{Frolov:Poly,BreTib1}. 

One of the conclusions of Refs.~\cite{Frolov:Exp,Frolov:Poly,BreTib1} is the existence of a mass gap to the formation of mini black holes if $f_s(-k^2) \sim k^2$ for large enough $k$. The presence of a mass gap in higher-derivative gravity models is known since the 1980s~\cite{Frolov:Weyl}, and it means that a black hole can only be formed if its mass is larger than a certain critical value. This contrasts with GR, where any mass can become a black hole, provided it is concentrated in a sufficiently small region.
On top of that, the curvature invariants remain bounded at $r = 0$ for all theories which $f_s(-k^2) \sim k^4$ (or faster) asymptotically, assuming that the collapsing shell has a finite thickness.


\subsection{Ultrarelativistic limit}
\label{ur-limit}

As a first step towards the gravitational field of a collapsing shell, we consider the field associated with an ultrarelativistic point-like particle, which can be obtained by the following procedure~\cite{Frolov:Exp,Frolov:Poly}. First, we perform a Lorentz transformation of the metric~\eq{isotr-metric} (with $h_i \equiv 0)$, 
\beq
\n{tbo}
t = \ga \left(t' - \be\, x' \right)
,
\qquad
x = \ga \left(x' - \be\, t' \right)
,
\qquad
\ga \equiv \frac{1}{\sqrt{1-\be^2}}, 
\eeq
which yields the metric of a moving object with velocity $\be$ in the $x$-direction. 
Then, we consider the Penrose limit,~\textit{i.e.}, we take $\, \be \to 1\,$ while keeping the relativistic mass 
\beq
\label{Pen}
M = \lim_{\ga \to \infty} ( \ga m ) 
\eeq
of the object fixed.

Therefore, after applying the boost~\eq{tbo}, one gets
\beq
\label{ds2}
\rd s^2 \,=\, \rd s_0^2 + \rd s_1^2
\,,
\eeq
where
\beq
\n{fuv}
\rd s_0^2 \,=\, - 2 \rd u \rd v + \rd y^2 + \rd z^2
\eeq
is the flat spacetime metric and
\beq
\rd  s_1^2 &=& - \frac{\ga^2\left(\ph+\psi\right)}{2}
\left[\left(1-\be\right)^2 \rd v^2  
+ \left(1+\be\right)^2 \rd u^2 
\right]
-\left(\ph-\psi\right) \, \rd u\, \rd v
\nonumber
\\
&&
-2 \psi \,(\rd y^2 + \rd z^2)
\,
\eeq
is the first-order perturbation. In the above formulas, we introduced the advanced and retarded null coordinates
\beq
v=t'+x', \qquad\qquad
u=t'-x'. 
\eeq

The form of the flat metric~\eq{fuv} remains unchanged in the limit $\be \to 1$, while the perturbation goes to
\beq
\label{gyr}
\rd  s_1^2 = {\Phi} \, \rd u^2 
\,, \qquad  \mbox{where}  \qquad
{\Phi} =  - 4 \, \lim_{\ga \to \infty}  ( \ga^2 \chi_2 ) 
\,.
\eeq
Hence, 
the dominant contribution in the ultrarelativistic limit comes from the spin-$2$ combination of the metric potentials, $\chi_2 = (\ph+\psi)/2$. This happens because, in this regime, 
the interaction between particles and the gravitational field is similar to that of photons (see, \textit{e.g.},~\cite{BuoGia,Giacchini:2018twk}).

The function ${\Phi}$ can be evaluated through~\eqref{gyr} by combining the heat kernel solution~\eqref{cHK} for $\chi_2$ and Eq.~\eqref{Pen}. 
Indeed, taking into account that after the boost $r^2 = \ga^2 u^2 + y^2 + z^2$, it follows
\beq 
\n{fizao}
{\Phi} =
-4 G \lim_{\ga \to \infty} ( \ga m )  
 \int_0^\infty \frac{\rd s}{s}\, h_2(s) \, e^{-(y^2+z^2) /4s} \lim_{\ga \to \infty} 
\frac{\ga e^{-\ga^2 u^2 /4s}}{\sqrt{4\pi s}}
\,.
\eeq
Recalling that
\beq
\lim_{\ga \to \infty} 
\frac{\ga e^{-\ga^2 u^2 /4s}}{\sqrt{4\pi s}}
\,=\, \de(u) \, ,
\eeq
the Eq.~\eq{fizao} can be written as
\beq 
{\Phi} \,=\,
-4 G M \, H(y^2+z^2) \, \de(u)
\,,
\eeq
where we define
\beq
\label{Fdiv}
H(\ze) \,\equiv\, \int_0^{\eta^2} \frac{\rd s}{s}\, h_2(s) \,e^{-\ze /4s}
\,.
\eeq
Notice that in~\eq{Fdiv} it was introduced an infrared cutoff $\eta$ for large $s$, because the integral in~\eq{fizao} typically has an infrared divergence, owing to the massless nature of the graviton. As discussed in~\cite{Frolov:Exp}, any change in the cutoff parameter can be absorbed into a redefinition of the coordinates. Observables, such as the curvature tensors, do not depend on $\eta$.

The metric~\eq{gyr} generalises, to higher-derivative gravity models, the Aichelburg--Sexl~\cite{Aich-Sexl} solution  in GR for the gravitational field of an ultrarelativistic massive particle without angular momentum (non-spinning gyraton). Furthermore, the extension of~\eq{gyr} to spinning objects can be found in~\cite{Boos:2020ccj} (see also~\cite{Kolar:2021uiu}). Finally, the solution~\eq{gyr} has been used in Ref.~\cite{hedao} to study the black hole formation by the head-on collision of ultrarelativistic particles, with similar conclusions about the existence of mass gap as 
the ones derived from the collapse of null shells --- which we mentioned at the beginning of the section and review in Sec.~\ref{egnonlocal} below.


\subsection{Thin null shell collapse}
\label{casca-fina}

Following~\cite{Frolov:Exp,Frolov:Poly}, we first consider a shell with vanishing thickness. At the linearised level, the field associated to a thin null shell, or $\delta$-shell, can be obtained by the continuous superposition of gyratons~\eqref{gyr} spherically distributed passing through one given point $O$~\cite{Frolov:Exp}. This point is the vertex of the null cone representing the shell, such that, for $t<0$, the shell collapses towards the apex $O$  and, for $t>0$, it proceeds its expansion after the collapse. The energy-momentum tensor $T_{\mu\nu}$ associated with the shell is 
\beq
T_{\mu\nu} \rd x^{\mu} \rd x^{\nu}={M\over 4\pi r^2}
[ \delta(v) \rd v^2 + \delta(u) \rd u^2]
.
\eeq

It can be shown that, outside the shell, the averaged metric perturbation $\langle \rd s_1^2\rangle$ resulting from this distribution of non-spinning gyratons is given by~\cite{Frolov:Exp}  
\beq
\label{dhr}
\langle \rd s_1^2\rangle  = - \dfrac{2GM }{ r} H(r^2-t^2) \bigg[ \bigg( \rd t-\frac{t}{ r} \rd r \bigg)^2 
+ \frac{r^2-t^2}{
2}\rd \Omega^2 \bigg] \,, \quad r \geqslant |t|\,,
\eeq
while in its interior (in the region where $r < |t|$) the spacetime is flat, $\langle \rd s_1^2\rangle = 0$. In~\eq{dhr}, $H(z)$ is defined by~\eqref{Fdiv}, as given by the metric~\eqref{gyr} associated with a single gyraton. 

The linearised Kretschmann 
scalar calculated with the metric~\eq{dhr} is 
\beq
&
\n{Kre-fino}
(R_{\mu\nu\al\be}^2)_{\rm lin} = \dfrac{48 G^2 M^2 \ze^2}{r^6} Q(\zeta),
&
\\ [2.5mm]
&
Q(\zeta) = 2 \left[ H''(\ze)\right] ^2 \zeta^2 + 2 H'(\ze) H''(\ze) \zeta + \left[ H'(\ze)\right] ^2
&
,
\eeq
where $\zeta = r^2 - t^2$. In this case,~\eq{Kre-fino} is singular at $r=0$ in any higher-derivative model~\cite{Frolov:Exp,Frolov:Poly,BreTib1}. However, this singularity is a consequence of the non-physical approximation of an infinitesimally thin shell.


\subsection{Thick null shell collapse}
\label{casca-grossa}

The metric associated with a thick null shell can be built, in the linear regime, by superposing a set of $\delta$-shells collapsing to the same spatial point $O$, which is taken as the origin of the coordinate system.  Suppose that the spherical collapsing null fluid is represented by a pulse 
with a finite time duration 
characterized by a density function $\rho(t)$. The total mass of the shell is 
\beq 
M = \int \rd t \, \rho(t),
\eeq
and it collapses with the speed of light, shrinking to zero radius  at the moment $t = 0$.
The corresponding metric perturbation can be obtained by averaging the metric~\eqref{dhr} of the thin null shells with respect to the density $\rho$~\cite{Frolov:Exp},
\beq
\n{llrr}
\langle \langle \rd s_1^2\rangle \rangle (t,r)
\,=\, 
\int \rd t' \, \rho(t') \langle \rd s_1 ^2 \rangle(t-t',r)
\,.
\eeq

Following~\cite{Frolov:Exp,Frolov:Poly}, we work with the simplest profile, assuming that the density $\rho(t)$ at $r=0$ remains constant during the collapse, and that it is null before and after it, namely
\beq
\n{rhodet}
\rho(t) = \left\{ 
\begin{array}{l l}
0 \, ,  &  \text{if $\,|t|>\tau/2\,$,}\\
M/\tau \, , &  \text{if $\,-\tau/2<t<\tau/2\,$,}\\
\end{array} \right .
\eeq
where $\tau$ is the thickness (or duration) of the shell.
The distribution~\eq{rhodet} defines specific spacetime domains; see the detailed discussion in~\cite{Frolov:Exp}. Here we restrict considerations to the $I$-domain (near the point $t = r = 0$), characterized by the intersection of the incoming and the outgoing fluxes of null fluid. Formally, it is defined as the locus of the spacetime points where $r + \vert t \vert < \tau/2$. 
In~\cite{Frolov:Exp,Frolov:Poly,BreTib1} it was proved that the non-singular source~\eq{rhodet} suffices to regularizes the Kretschmann scalar for theories with form factors such that $f_s (-k^2) \sim k^4$ or faster for large $k$. 

Taking into account that only the $\delta$-layers which cross $O$ at times $t^\prime \in (t-r,t+r)$ contribute to the field inside the $I$-domain, it is not difficult to verify that the metric is stationary there and that the Eq.~\eq{llrr} yields~\cite{Frolov:Exp}
\beq
\label{Met_Thick}
\hspace{-1cm}
\langle \langle \rd s_1^2\rangle \rangle
= -\frac{2G M}{\tau r}\left[  c_0 \,  \rd t^2 + c_2 \, \frac{\rd r^2}{r^2}
+  \frac{1}{2} \left( c_0 r^2  - c_2 \right) \rd \Omega^2
\right] ,
\quad r + \vert t \vert < \frac{\tau}{2} \, ,
\eeq
where 
\beq
\label{J-def}
c_n(r) \equiv \int_{-r}^{r} \rd \xi \, \xi^n\  H (r^2-\xi^2) 
\eeq
and the function $H(z)$ is defined in~\eqref{Fdiv}. 


\subsection{Example: Gaussian form factor}
\label{egnonlocal}

As an explicit example of the procedure outlined in this section, here we consider the case of the form factor~\eq{exp-form factor}, namely, $f_s(-k^2)=\exp(k^2/\mu_s^2)$, which was studied in~\cite{Frolov:Exp}. The first step is to obtain the inverse Laplace transform $h_2 (s)$, defined in~\eq{H-1}, and then evaluate the function in~\eq{Fdiv}. It is not difficult to show that, in this case, $h_2 (s) = - \theta (s - \mu_2^{-2})$ ($\theta(x)$ is the Heaviside step function) and 
\begin{equation}
\n{H-nl}
H(\zeta) = \ln (\ze/\eta^2) + \ga_{\rm E} +  E_1 ( \mu_2^2 \ze /4),
\end{equation}
where $\ga_{\rm E}$ is the Euler--Mascheroni constant and $E_1 (x)$ is the exponential integral function.

Using \eq{H-nl}, we get for~\eq{Kre-fino}
\beq
(R_{\mu\nu\al\be}^2)_{\rm lin} = \frac{3 G^2 M^2 \mu_2^4  \ze^2}{r^6} + O\left(\ze^3\right)
,
\eeq
showing that as the collapse of the thin null  shell proceeds, the Kretschmann scalar diverges for $r \to 0$. On the other hand, if the shell has some thickness, the curvature gets regularized. Indeed, the Eqs.~\eq{J-def} and~\eq{H-nl} imply the small-$r$ behaviour for the components of the metric~\eq{Met_Thick},
\begin{subequations} \n{CJ}
\begin{align} 
c_0=&\,\,{ \mu_2^{2} \, r^3 (420 - 21 \mu_2^{2} r^2+ \mu_2^4 r^4  )\over 1260 }+\ldots\, ,\\[2.5mm]
c_2=&\,\,{\mu_2^2\, r^5(756 -27\mu_2^2 r^2+\mu_2^4 r^4)\over 11340}+\ldots\,,
\end{align}
\end{subequations}
so that we get for the Kretschmann invariant in the $I$-domain,
\beq
\lim_{r \to 0} \, (R_{\mu\nu\al\be}^2)_{\rm lin} = {32 G^2 M^2 \mu^4_2 \over 3 \tau^2}.
\eeq
This shows that the curvature remains finite at $r=0$ for the thick shell model.

Finally, let us discuss the mini black hole formation, which is 
related to the invariant 
\beq
(\nabla r)^2={1\over 4 \varsigma} \, g^{\mu\nu} \na_\mu \varsigma \na_\nu \varsigma, \qquad
\varsigma  = g_{\theta\theta}\, . 
\eeq
A curve in the $tr$-plane such that $(\nabla r)^2 = 0$ is an apparent horizon. 
Using~\eq{Met_Thick} and~\eq{CJ} one finds that, to the leading order in $M$, this invariant in the $I$-domain is given by 
\beq
(\nabla r)^2=1-{2M \mu^2_2 r^2\over  \tau}.
\eeq
Since in this domain $r<\tau$, we have
\beq
(\nabla r)^2> 1-{2M\mu^2_2 \tau}\, .
\eeq
This relation means that, for a given $\mu_2$ and a fixed duration $\tau$ of the pulse, the mini black hole does not form
if the mass $M$ is small enough~\cite{Frolov:Exp}.


\section{Towards the non-linear regime}
\label{sec:non-linear}

The presence of higher-derivative terms in the gravitational action makes the analysis of the complete non-linear equations of motion a highly difficult task, and non-locality poses an extra obstacle, because of the infinite-order derivatives.
In particular, differently from the linear limit [where any higher-derivative structure could be reduced to the form of the action~\eqref{quad-action-reduced}], the space of solutions of the complete field equations depends on the terms that are actually in the action.

For example, if the Einstein--Hilbert action is enlarged only by the quadratic structures $R {F}_1(\Box) R$ and/or $R_{\mu\nu} {F}_2(\Box) R^{\mu\nu}$, the Ricci-flat solutions of GR are also vacuum solutions in the non-local model; however, if the term quadratic in the Riemann tensor is included, then those Ricci-flat spacetimes are not solutions of non-local gravity~\cite{Li:2015bqa}. The stability of these type of solutions, imported from GR, was studied in~\cite{Myung:2017qtc,Calcagni:2017sov,Briscese:2019rii}, and in~\cite{Calcagni:2018pro} for theories constructed with form factors that are functions of the Lichnerowicz operator; thermodynamic aspects of these black holes were also discussed in~\cite{Conroy:2015wfa,Myung:2017axf,Xiao:2021maa,Xiao:2021ewv}.

It is worth noticing that this process of constructing solutions from the comparison with GR  is not valid in the whole spacetime, namely, it might not correctly reflect the interaction between gravity and the matter source. For instance, while the static Schwarzschild solution in GR can be associated with a Dirac delta source (see, \textit{e.g.},~\cite{Balasin:1993fn}), non-locality and higher derivatives induce a smearing of the delta source at linear level~\cite{BreTib2,Jens,Nos6der,Modesto:2011kw,Tseytlin95,Li:2015bqa,hedao,Koshelev:2018hpt} (see also Sec.~\ref{SecPon} above), and their effect might be able to regularize the singularity.

Since exact static black hole solutions in non-local gravity with matter sources are still an open problem, some insights were obtained from the linearised limit (as discussed above, in this chapter) and also from a non-linear approximation of the field equations, which we describe in what follows. To this end, let us consider the models of the type~\eqref{only-spin-2-action}, {\it i.e.}, with form factors $F_2(\Box) = -2 F_1(\Box)$, so that $f_0(-k^2) = f_2(-k^2) = e^{\ga(\Box)}$. Because of this specific form of the higher-derivative sector, the form factor gets factored together with the Einstein tensor in the field equations~\cite{Modesto:2011kw,Zhang14}, namely
\beq
\n{fefr}
e^{\ga(\Box)} G^{\mu} {}_{\nu} + {O}({R}^2_{\ldots}) = 8 \pi G \, T^{\mu} {}_{\nu}
.
\eeq 
In the works~\cite{Modesto:2011kw,Modesto:2012ys,Zhang14,Bambi:2013gva,Bambi:2016uda} the approximated version of this equation was considered, in the form
\beq
\label{Gff}
G^{\mu} {}_{\nu} = 8 \pi G \, \tilde{T}^{\mu} {}_{\nu},
\eeq
where 
\beq
\n{Tmn-eff-0}
\tilde{T}^{\mu} {}_{\nu} = e^{-\ga(\Box)} {T}^{\mu} {}_{\nu}
\eeq 
is an effective energy-momentum tensor. 

The Eq.~\eq{Gff} can be regarded as an approximation of~\eq{fefr} by discarding the terms ${O}({R}^2_{\ldots})$ of higher order in curvatures. Furthermore, taking the d'Alembertian in~\eq{Tmn-eff-0} in its flat-spacetime form, for a point-like source $T_{00} = m \de^{(3)}(\vec{r})$, we get
\beq
\n{T00-eff}
\tilde{T}_{00} = \rho_{\rm eff}, \qquad {\rm where} \qquad \rho_{\rm eff} (r) = \frac{m}{2 \pi^2 r} \int_0^\infty \rd k \, \frac{k \sin(kr)}{e^{\ga(-k^2)}} ,
\eeq 
just like the expression~\eq{lap-eff}. Alternatively, the truncations in the original equations~\eq{fefr} and~\eq{T00-eff} might be compensated by imposing the conservation of the effective energy-momentum tensor, \textit{i.e.}, $\nabla_\mu \tilde{T}^\mu {}_\nu = 0$~\cite{Modesto:2011kw}. This leads to the introduction of the effective radial ($p_r$) and tangential ($p_\th$) pressures in the energy-momentum tensor
\beq
\n{effT}
\tilde{T}^{\mu} {}_\nu = \text{diag}(-\rho_{\rm eff} , p_r , p_\th , p_\th ) .
\eeq

In this spirit, generalisations of the Schwarzschild solution were obtained by using the metric \textit{Ansatz}
\beq
\label{metricAnsatz}
\rd s^2 = - A(r)  \rd t^2 + \frac{\rd r^2}{A(r)} + r^2 \rd \Omega^2.
\eeq
Then, making the redefinition
\beq
A(r) = 1 - \frac{2G \tilde{m}(r)}{r}
\eeq
and substituting the above expressions into~\eq{Gff}, one obtains the equations
\beq
 \n{EqGtt}
\frac{\rd \tilde{m}}{\rd r} & = & 4 \pi r^2 \rho_{\rm eff} \,=\, - 4 \pi r^2 p_r
\,, 
\\
 \n{Conserva}
\frac{\rd p_r}{\rd r}  & = &  \frac{2}{r} \left( p_\th - p_r \right) - \frac{\left( p_r + \rho_{\rm eff} \right)}{2A} \frac{\rd A}{\rd r} ,
\eeq
which are solved by 
\beq
p_r = - \rho_{\rm eff}, \qquad p_\th = - \rho_{\rm eff} - \frac{r \rho_{\rm eff}^\prime}{2}
\eeq
and
\beq
\n{meffinal}
\tilde{m}(r) = 4\pi \int_0^r \rd x \, x^2 \rho_{\rm eff} (x) 
.
\eeq
Notice that $\tilde{m}(r)$ is the same mass function as the one defined in~\eq{massfunction}.

\begin{figure}[t]
\begin{center}
\includegraphics[scale=0.6]{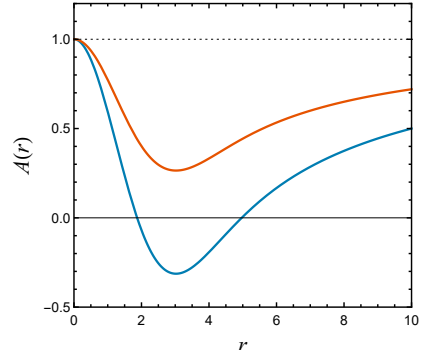}
\caption{Plot of $A(r)$ 
for $\mu = M_{\rm P} = 1$ in two situations. In blue line, with $m = 2.5$ (larger than the critical mass), the metric has two horizons at $r_- = 1.87$ and $r_+ = 4.67$. In red line, with $m = 1.4$ (smaller than the critical mass), the metric has no horizon.}
\label{fig:8}     
\end{center}
\end{figure}

For the Gaussian form factor~\eq{exp-form factor} the solution is the same as the one obtained in~\cite{Modesto:2010uh,Nicolini:2005vd}, namely,
\beq
\label{NSSbh}
\rd s^2 = - \left[ 1 - \frac{4Gm}{r \sqrt{\pi}} \ga\left( \tfrac{3}{2} , \tfrac{\mu^2 r^2}{4} \right)  \right]   \rd t^2 +  \frac{\rd r^2}{1 - \frac{4Gm}{r \sqrt{\pi}} \ga\left( \tfrac{3}{2} , \tfrac{\mu^2 r^2}{4} \right)} + r^2 \rd \Omega^2 
,
\eeq
where  $\ga\left( a , x \right)$ is the lower incomplete gamma function, or in terms of the error function,
\beq
\ga\left( \tfrac{3}{2} , \tfrac{\mu^2 r^2}{4} \right) 
= \frac{1}{2}  \left[ \sqrt{\pi} \, {\rm erf} \left(\frac{\mu r}{2}\right) - \mu r \, e^{-\mu^2 r^2/4} \right] .
\eeq
Like the case discussed in Sec.~\ref{egnonlocal}, there exists a mass gap for this solution to describe a black hole. In fact, if 
\beq
m \, <  \, 1.9 \, \frac{M_\text{P}^2}{\mu},
\eeq
(we denote $1/G = M_\text{P}^2$) the invariant $(\na r)^2 = A(r)$ is always positive and the metric does not have any horizon~\cite{Nicolini:2005vd}. On the other hand, for values of $m$ larger than this critical mass, the solution displays two horizons. These features are shown in Fig.~\ref{fig:8}, where we plot the graph of $A(r)$ for the two scenarios.

In what concerns the regularity of the solution, it is straightforward to verify that the metric~\eq{NSSbh} is regular, as well as the related curvature invariants~\cite{Nicolini:2005vd,Modesto:2010uh}. This is expected, for the metric has a de~Sitter core, 
\beq
A(r)  = 1 + \frac{G m \mu^3 r^2}{3 \sqrt{\pi }} + {O}(r^4).
\eeq
Moreover, since the components of~\eq{NSSbh} are even analytic functions of $r$, it follows that all its local curvature-derivative invariants are finite~\cite{Giacchini:2021pmr,Nos6der}.

Similar solutions were obtained in the cases of more general exponential form factors~\eq{exp-generic-form factor} of the type $f(-k^2)=\exp (k^2/\mu)^n$, and it was shown that for larger $n$ the solutions can have more than two horizons~\cite{Zhang14}. The possibility of having multi-horizon black holes also occurs in the case of the Kuz'min--Tomboulis form factor~\eq{kuzm-form factor}, as numerically shown in~\cite{Modesto:2011kw,Zhang14}.


\section{Concluding remarks}
\label{sec:conclus}

In this chapter, we 
reviewed linearised metric solutions in both stationary and dynamical scenarios. In the former case, we analysed 
static and rotating spacetimes and showed that non-locality can regularize point-like and ring-like singularities. 
In the latter scenario, we showed that regularized mini black hole solutions can be found and that the formation of an apparent horizon can be avoided as long as the mass of the object is smaller than some critical value set by the scale of non-locality. Furthermore, we discussed an attempt towards finding full non-linear solutions 
and showed that regular solutions can be obtained 
by working with some simplified field equations.

Understanding the physics of non-locality at the full non-linear level remains one of the most outstanding open questions in the context of ghost-free non-local theories of gravity. In fact, it is still not entirely clear whether singularities can be really avoided in the non-linear regime or whether a horizon can form. Indeed, it has also been argued that non-locality could prevent the formation of a horizon, for any value of the mass, in such a way that black holes could be replaced by ultra-compact horizonless objects in some non-local gravity models~\cite{Koshelev:2017bxd,Buoninfante:2019swn}, although a rigorous proof of this statement is still lacking.

Providing definite answers to these questions is challenging but at the same time very important and stimulating. Future investigations and new ideas to solve infinite-derivative non-linear differential equations are surely needed. In fact, finding a full non-linear black hole-like solution can place non-local gravity on firmer ground and lay the foundation for future phenomenological studies in astrophysics, \textit{e.g.}, in the context of binary mergers and gravitational waves, and thus offer a new scenario where non-local extensions of Einstein's GR can be tested and constrained.


\begin{acknowledgement}
Nordita is supported in part by NordForsk.
\end{acknowledgement}



\end{document}